\documentstyle[12pt]{article}
\newcommand{\bq}{\begin{equation}}
\newcommand{\ee}{\end{equation}}
\newcommand{\al}{\mbox{$\alpha$}}
\newcommand{\lm}{\mbox{$\log(1/\alpha)$}}
\newcommand{\r}{\mbox{$\vec{r}$}}
\newcommand{\pv}{\mbox{$\vec{p}$}}
\newcommand{\bi}[1]{\bibitem{#1}}
\newcommand{\fr}[2]{\frac{#1}{#2}}

\begin{document}
\normalsize
\begin{center}
{\Large \bf Budker Institute of Nuclear Physics}
\end{center}
\begin{flushright}
BINP 94-30\\
March 1993
\end{flushright}
\vspace{1.0cm}
\begin{center}{\Large \bf Large relativistic corrections\\
to the positronium decay rate}\\
\vspace{1.0cm}
{\bf I.B. Khriplovich}\footnote{e-mail address: KHRIPLOVICH@INP.NSK.SU}
and {\bf A.I. Milstein}\footnote{e-mail address: MILSTEIN@INP.NSK.SU}\\
Budker Institute of Nuclear Physics, 630090 Novosibirsk, Russia\\
and Novosibirsk State University
\vspace{4.0cm}
\end{center}

\begin{abstract}
Relativistic corrections to the positronium decay rate are calculated.
They are close to $40(\al /\pi)^2$ and $46(\al /\pi)^2$ for
singlet and triplet states respectively.
\end{abstract}

\newpage
1. The strong disagreement between the experimental value of the
orthopositronium decay rate\cite{NGR}
\bq\label{eq:Gex}
\Gamma_{exp}^{o-Ps}=7.0482\pm 16\; \mu s^{-1}.
\ee
and its theoretical value which includes the order $\al$ and $\al^2 \lm$
corrections\cite{OP,CLS,CasLep,Adk}
\bq \label{eq:Gt}
\Gamma_{th}^{o-Ps}=m\al^{6}\fr{2(\pi^{2}-9)}{9\pi}\left[1-10.28\fr{\al}{\pi}-
\fr{1}{3}\al^{2}\log\fr{1}{\al} \right]=7.03830\; \mu s^{-1}.
\ee
is a serious challenge to modern QED. For the disagreement to be resolved
within the QED framework the correction $\sim (\al/\pi)^{2}$, which has not
been calculated up to now, should enter the theoretical result (\ref{eq:Gt})
with the numerical factor $250\pm 40$ which may look unreasonably large.

Such a hope is not as unreasonable however.  The argument is as
follows\cite{KY}. The large, $-10.28$, factor at the $\al/\pi$ correction
to the decay rate (see (\ref{eq:Gt})) means that the typical magnitude of
the factor at the $\al/\pi$ correction to the decay amplitude is roughly
5. Correspondingly, this correction squared contributes about
$25(\al/\pi)^{2}$ to the decay rate. Indeed, numerical calculations
\cite{Bur} have given factor $28.8\pm 0.2$ at $(\alpha/\pi)^2$ in this
contribution.

Moreover, it is only natural to expect that the interference of the
second-order radiative correction to the amplitude with the zeroth-order
amplitude should contribute about twice as much to the decay rate as the
square of the first-order correction. In other words, the natural scale
for the total second-order radiative correction to the decay rate can
be about\cite{KMY}
\begin{equation}\label{eq:sq}
100(\alpha/\pi)^{2}.
\end{equation}

A similar conclusion is made in a recent paper\cite{SL} starting from the
Pade approximants.

One more class of large contributions to the positronium decay rate is
relativistic corrections. A simple argument in their favour is that the
corresponding parameter $(v/c)^2 \sim \al^2$ is not suppressed, as
distinct from that of usual second-order radiative corrections,
$(\al/\pi)^2$, by the small factor $1/\pi^2 \sim 1/10$. In this article we
present the results of calculations of relativistic corrections to the
positronium decay rate.

This problem was addressed previously in Refs.\cite{KU,LLM}. We differ
essentially from those authors in the approach to the problem and, which
is more essential, in the conclusions made. The origin of the
disagreements will be elucidated below.

As to the relativistic correction to the parapositronium decay rate, also
obtained in the paper, its calculation was started by us as a warm-up
exercise for the much more complicated orthopositronium problem. However,
the correction in the singlet case also turns out large, quite close to
the sensitivity of the recent experiment\cite{AG}.

2. The central point when treating the relativistic corrections to the
positronium decay rate is as follows. Calculating the decay amplitude we
have to integrate the annihilation kernel over the distribution of the
electron and positron three-momenta $\vec{p}$. To lowest approximation in
$v/c$ the kernel, both for para- and orthopositronium, is independent of
those momenta and we are left with the integral over $\vec{p}$ of the
nonrelativistic wave function in the momentum representation which is
equivalent to $\psi (r=0)$ in the coordinate one. However, already to
first order in $(p/m)^2$ the momentum integral
\bq
\int d\vec{p}\, (p/m)^2\psi(\vec{p})\, =\,
\int d\vec{p}\, (p/m)^2 \fr{8\sqrt{\pi a^3}}{(p^2 a^2 +1)^2}
\ee
linearly diverges at $p\rightarrow\infty$ ($a=2/m\al$ is
the positronium Bohr radius). Crucial for the problem is the following
observation. The true relativistic expression for the annihilation kernel
does not grow up at $p\rightarrow\infty$, as distinct from its expansion
in $p/m$. So, its integral with $\psi (p)$ in fact converges. Let us
transform therefore the integral over $|\vec{p}|$ into that from
$-\infty$ to $+\infty$ and shift the integration contour into the upper
halfplane. We will first come across the wave function pole at $p=i
m\al/2$ and then the relativistic branching point at $p=i m$ which
originates from the amplitude and is not related by itself to the wave
function. It is obviously just the pole contribution which corresponds to
the relativistic correction we are looking for. This contribution can be
easily calculated and constitutes
$$-\frac{3}{4}\al^2 \psi (r=0)$$.
In other words, the recipe for treating the relativistic corrections
originating from the amplitude is just to make the substitution
\bq\label{eq:c}
(p/m)^2=v^2\rightarrow -\frac{3}{4}\al^2
\ee
in them. One may wonder about the sign in rhs of this relation. Let us
have in mind however that the main contribution to the integral comes from
the relativistic cut. That contribution corresponds to usual radiative
corrections $\sim \al/\pi$ and is of course much larger than the effect
$\sim \al^2$ we are interested in.

The next point essential for our consideration is the use of noncovariant
perturbation theory (see, e.g., Ref.\cite{KMY1}) which allows us to treat
in a natural way the positronium binding energy.

Let us start with a more simple case of parapositronium. Here the
noncovariant annihilation amplitude can be written as
\begin{eqnarray}
&\displaystyle
M=4\pi\al V^{+}(\vec{e}_2 \vec{\al})\frac{\Lambda_{+}(\vec{p}-\vec{k}_1)
-\Lambda_{-}(\vec{p}-\vec{k}_1)}
{E-\omega-\epsilon(\vec{p}-\vec{k}_1)-\epsilon(p)}(\vec{e}_1 \vec{\al})U
 + (1\leftrightarrow 2)\, ,\nonumber\\
&\displaystyle
V=\sqrt{\fr{\epsilon (p)+m}{2\epsilon (p)}}\left(1-
\fr{\vec{\al}\pv}{\epsilon (p)+m}\right)
\left(\begin{array}{c}0\\ \chi\end{array}\right)\, ,
\, U=\sqrt{\fr{\epsilon(p)+m}{2\epsilon(p)}}\left(1+\fr{\vec{\al}\pv}
{\epsilon(p)+m}\right)\left(\begin{array}{c}\phi\\ 0\end{array}\right)\, .
\end{eqnarray}
In this expression $\chi$ and $\phi$ are nonrelativistic spinors;
$E=2m-m\al^2/4$ is the positronium total energy;
$\vec{e}_{1,2}$ and $\vec{k}_{1,2}$ are the polarizations and momenta of
the photons; $\omega_1=\omega_2=\omega=E/2$ are their frequencies;
$\epsilon(p)=\sqrt{m^{2}+p^{2}}$;
$$\Lambda_{\pm}(\vec{p})=\frac{1}{2}\left( 1\pm\frac{\vec{\alpha}\pv+
\beta m}{\epsilon(p)}\right)$$
are the projectors onto the positive and negative energy
states of a fermion with a momentum \pv\ correspondingly. The Coulomb
interaction in the intermediate state can be neglected since the momentum
of one particle in it is close to $m$.

The expansion of the amplitude in $p/m$ is straightforward. Averaging over
the directions of $\vec p$ (an $S$-state is under discussion) and using
relation (\ref{eq:c}) we obtain
\bq\label{eq:s1}
M+\delta M\, =\, [1+\al^2(\frac{1}{2}+\frac{\sqrt{2}}{8})]M
\ee
where $M$ is the lowest order annihilation amplitude.
The corresponding relative correction to the decay rate is
\bq\label{eq:s2}
\frac{\delta\Gamma}{\Gamma}=\al^2(1+\frac{\sqrt{2}}{4})=1.35\al^2.
\ee

3. The calculation of relativistic corrections for the triplet state
decaying into three photons is much more tedious problem. We believe that
have managed to simplify it considerably, but still it is too lengthy to
be presented in detail in a short letter. So, only its brief outline is
given below.

The construction of noncovariant perturbation theory amplitude is
straightforward. But it is convenient to treat it in a slightly different
way than it was done for the singlet state. First of all rewrite the
initial energy $E$ in the perturbative denominators as
$$E=E-2\epsilon(p)+2\epsilon(p)$$
and expand the amplitude in
$$E-2\epsilon(p)=\frac{m\al^2}{2}\, $$
(here we use recipe  (\ref{eq:c}) for $p^2/m^2$).

Zeroth term of the expansion transforms into usual covariant Feynman
amplitude for electron and positron with 4-momenta $(\epsilon (p),
\pm\vec{p})$. Now we expand this ``covariant'' amplitude in $p/m$, average
the terms of second order in $p/m$ over the directions
of $\vec{p}$ and make the substitution (\ref{eq:c}). The interference of
this $\al^2$-correction with the lowest order amplitude after the
summation over the photons polarizations and integration over the final
phase space generates the following correction to the decay rate:
\bq\label{eq:cov}
\frac{\delta\Gamma_c}{\Gamma}=\al^2\frac{31\pi^2-240}{16(\pi^2-9)}.
\ee

This correction is conveniently combined with that originating from the
phase space correction. It can be easily demonstrated that the shift of
the total energy from $2m$ to $E=2m-m\al^2/4$ changes the phase space and
therefore the decay rate by
\bq
\frac{\delta\Gamma_p}{\Gamma}=-\al^2\frac{1}{4}.
\ee

In this way we come to the following total ``covariant'' correction to the
decay rate
\bq\label{eq:s}
\frac{\delta\Gamma_c+\delta\Gamma_p}{\Gamma}
=\al^2\frac{27\pi^2-204}{16(\pi^2-9)}.
\ee
We have checked that if one goes over from $\al^2$ to $v^2$ in
Eq.(\ref{eq:cov}) according to prescription (\ref{eq:c}) and changes the
phase space shifting $2m\rightarrow 2m+mv^2$, the result coincides with
the $v^2$ correction to the probability of $3\gamma$ annihilation of
free electron and positron in $^3S_1$ state calculated in Ref.\cite{KU}
(see also Ref.\cite{LLM}).

The correction to the decay rate induced by the term of first order in
$E-2\epsilon(p)=m\al^2/2$ in the expansion of exact amplitude
demands numerical calculations which give
\bq\label{eq:ncov}
\frac{\delta\Gamma_n}{\Gamma}=0.807\al^2.
\ee
Let us note here that the weird term with $\sqrt{2}$ in the correction to
the singlet decay rate (see Eqs. (\ref{eq:s1}), (\ref{eq:s2})) is of the
same ``noncovariant'' origin.

Going back  to the triplet decay rate, we have to note that our results
for the $\al^2$ corrections themselves differ completely from those of
Refs.\cite{KU,LLM}. It is not so much due to the ``noncovariant''
correction (\ref{eq:ncov}) completely lost there, this correction is not
so large
numerically. The main problem is that of translating $v^2$ into $\al^2$.
In Ref.\cite{KU} they use the prescription $v^2\rightarrow\al^2$ which
leads to a wrong sign of the $\al^2$ correction (though to a reasonable
absolute value). On the other hand, the prescription $v^2\rightarrow
-\al^2/4$ used in fact in Ref.\cite{LLM} gives a correct sign, but
strongly underestimate the effect. Quite possibly however, there is no
direct contradiction between their result and ours, since as it is stated
explicitly in Ref.\cite{LLM}, their consideration refers to a part of
relativistic corrections only.

4. Let us consider at last the effects originating from relativistic
corrections to the wave function $\psi(\vec{r})$ itself. We will use here
the Breit equation following to some extent Ref.\cite{KY}. Para- and
orthopositronium can be treated here in parallel.

The part of the Breit Hamiltonian (BH) that corresponds to the relativistic
corrections to the dispersion law of the particles and to their Coulomb
interaction,
\bq
V_{c}=-\fr{p^{4}}{4m^{3}}+\fr{\pi\al}{m^{2}}\delta(\r)\,,
\ee
can be easily transformed to
\bq\label{eq:co}
V_{c} = \fr{\al^{3}}{8r}.
\ee
Here and below we omit constant terms in the perturbations (obviously,
they do not change the wave function) and substitute $-m\al/2$ for
$\partial_{r}$ acting on the ground state positronium wave function.

The next spin-independent term in the BH
\bq
V_{m}=-\fr{\al}{2m^{2}r}\left(p^{2}+\fr{1}{r^{2}}\r(\r\pv)\pv\right)\,,
\ee
describes the magnetic electron-positron interaction due to the orbital
motion. For the ground state it transforms into
\bq\label{eq:m}
V_{m} = \fr{\al^3}{4r}-\fr{\al^{2}}{2mr^{2}}\,.
\ee

The last term in BH of interest for our problem is the
contact spin-spin interaction
\bq\label{eq:V3}
V_{ss}=\fr{\pi\al}{m^{2}}A\delta(\r)\,;\;\; A=\fr{7}{3}S(S+1)-2.
\ee
It is conveniently rewritten as
\bq\label{eq:ss}
V_{ss}=A\fr{1}{4m}\left[H,\fr{\al}{r}\right] + A\fr{\al^2}{4mr^2};\;\;
H=p^2/m-\al/r.
\ee

The terms $\al^3/8r,\;\;\al^3/4r$ from Eqs. (\ref{eq:co}),
(\ref{eq:m}) taken together shift obviously the coupling
constant $\al\rightarrow \al(1-3\al^2/8)$ which leads to the
following relative correction both to the $|\psi(0)|^2$ and decay rate
\bq
\fr{\delta\Gamma_1}{\Gamma}=-\fr{9\al^2}{8}.
\ee

As easily one can calculate the relative correction due to the commutator
term in Eq. (\ref{eq:ss}):
\bq
\fr{\delta\Gamma_2}{\Gamma}=A\fr{\al^2}{2}.
\ee

Let us turn now to the singular part of the Breit perturbation
\bq
V_2=\fr{\lambda}{mr^2};\;\;\lambda=\,\al^2(\fr{A}{4}-\fr{1}{2}).
\ee
The normalized solution of the radial wave equation
\bq
\left(\fr{1}{r}\fr{d^2}{r^2}r -\fr{\lambda}{r^2}
+\fr{m\al}{r}+m\tilde{E}\right)R=0
\ee
is
\bq
R=2(m\al/2)^{3/2}[1-\lambda(3-C)](m\al r)^{\lambda}
\exp {[-(1-\lambda)m\al r/2]}
\ee
where $C=0.577$ is the Euler constant. The eigenvalue $\tilde{E}$
deviation from $-m\al^2/4$ is by itself irrelevant to our problem.
The corresponding relative correction to the $|\psi(0)|^2$ and decay rate
constitutes obviously
\bq
\fr{\delta\Gamma_3}{\Gamma}=-2\lambda[(3-C)-\log(m\al r_0)]
\ee
where $r_0\sim 1/m$ is the distance at which the annihilation takes place.
The logarithmically enhanced part of this correction
\bq
\al^{2}\lm\left\{\begin{array}{cc}
                            2,   &  S=0 \\ -1/3,  &  S=1  \end{array}
\right.\,.
\ee
has been calculated previously for triplet (see formula (\ref{eq:Gt}) and
singlet cases in Refs.\cite{CasLep} and \cite{KY} respectively. So, we
omit it and in this way come to the following total relativistic
correction due to the $\psi$-function modification:
\bq
\fr{\delta\Gamma_{\psi}}{\Gamma}=\al^{2}\left\{\begin{array}{cc}
                  31/8-2C-2\log(mr_0),   &  S=0 \\
                 -19/24+1/3C+1/3\log(mr_0),  &  S=1  \end{array}
\right.\,.
\ee

We believe that $\pm 1$ is a fair estimate for the scatter of possible
values of $\log(mr_0)$ introduced by the uncertainty in the short-distance
cut-off $r_0$. On the other hand, the cut-off of the logarithmic
contribution at the atomic distances has been taken care exactly in our
consideration.

Our result for the atomic relativistic correction in orthopositronium,
$(-19/24+1/3C)\al^2=-0.6\al^2$, differs from that given in Ref.\cite{LLM}. We
cannot explain the disagreement, since the authors of Ref.\cite{LLM} present
only their numerical result for this correction: $1.16\al^2$.

5. To summarize, the total relativistic corrections in para- and
orthopositronium constitute, respectively,
\bq
\fr{\delta\Gamma}{\Gamma}=4.1\al^2=40(\fr{\al}{\pi})^2,\;\;S=0;
\ee
\bq\label{eq:res}
\fr{\delta\Gamma}{\Gamma}=4.7\al^2=46(\fr{\al}{\pi})^2,\;\;S=1
\ee
(it is instructive perhaps to present these relativistic
corrections in usual ``radiative'' units $(\al/\pi)$ as well).
The terms with $\log(mr_0)$, omitted here, introduce the errors which we
estimate as
\bq
\pm 2\al^2=\pm 20(\fr{\al}{\pi})^2,\;\;S=0;
\ee
\bq
\pm \fr{1}{3}\al^2=\pm 3(\fr{\al}{\pi})^2,\;\;S=1.
\ee

As to the orthopositronium decay rate, our correction (\ref{eq:res}) and
that of Ref.\cite{Bur}, taken together, reduce essentially the gap between the
theory and experiment, from $(250\pm 40)(\al/\pi)^2$ to $(175\pm 40)
(\al/\pi)^2$.

In parapositronium the magnitude of the calculated correction is close to
the present experimental accuracy\cite{AG}. Here there are no special
reasons to expect that true radiative corrections are as large. So the
measurement of the effect looks sufficiently realistic.

\bigskip
We are extremely grateful to A.S. Yelkhovsky for numerous useful
discussions and the participation in some stages of the work. We
acknowledge the financial support by the Program ``Universities of
Russia'', Grant No.94-6.7-2053. One of us (I.B. Kh.) thanks the Cambridge
University for the kind hospitality and the SERC for financial support.

\end{document}